# The Role of Ionization in Thermal Transport of Solid Polyelectrolytes

Xingfei Wei[1] and Tengfei Luo[*, 1, 2, 3]

[1]Department of Aerospace and Mechanical Engineering, [2]Department of Chemical and Biomolecular Engineering, [3]Center for Sustainable Energy at Notre Dame (ND Energy), University of Notre Dame, Notre Dame IN 46556, United States

* Corresponding author: E-mail: tluo@nd.edu, (574) 631-9683.

## ABSTRACT

Amorphous polymers are known as thermal insulators, increasing their thermal conductivities have not been guided by fully understood physics. In this work, we use molecular dynamics simulations to study the thermal transport mechanism of solid polyelectrolytes, poly(acrylic acid) (PAA) and its ionized forms. The thermal conductivity of PAA increases monotonically with the ionization strength. Although stronger ionization induces larger Coulombic interactions, the Coulombic interaction does not directly contribute to the thermal conductivity enhancement. Instead, it enhances thermal transport through the Lennard-Jones (LJ) interaction. The strong Coulombic force between the counterion and the ionized carboxylic group shifts the LJ force to the stronger LJ repulsive regime, which is mainly responsible for the improved thermal conductivity. Applying a high pressure can further reduce the inter-atomic distance and trigger the thermal transport through the LJ interaction. A thermal conductivity of 1.09 W/m.K can be achieved at 11.2 GPa in an ionized PAA.

# 1. INTRODUCTION

Amorphous polymers are known to have low thermal conductivity in the range of 0.1-0.5 W/m.K.[1] The search for thermally conductive polymers is of tremendous scientific interests and practical application significance.[2, 3] While there have been intensive efforts to improve polymer thermal conductivity through compositing with inorganic fillers, the overall thermal conductivity is still largely limited by the low thermal conductivity of polymer matrices.[4] It has been pointed out that a polymer matrix with a thermal conductivity >1 W/m.K will makes compositing more meaningful, e.g., compositing such a polymer with graphite can lead to a thermal conductivity >20 W/m.K, which is larger than stainless steel (~16 W/m.K).[1, 2, 5]

Changing the global morphology of polymer from amorphous to crystalline has been proven effective in improving thermal conductivity by orders of magnitude from O(0.1) to O(10) W/m.K,[6] highlighting the importance of molecular chain conformation in thermal transport.[7-9] However, these materials have highly anisotropic thermal conductivity, limiting their applications as bulk materials. Recently, there have been a few studies reporting amorphous polymers with thermal conductivity greater than 1 W/m.K,[10-13] and most of them attribute the high values to the microscopic chain conformation. It is found that when the polymer chains in the amorphous bulk phase has greater spatial extension, usually characterized by radius of gyration ($R_g$), the thermal transport along the covalent backbone becomes more effective, leading to higher thermal conductivity.[11, 14, 15] In 2017, a report shows that an electrostatically engineered polyelectrolyte, ionized poly(acrylic acid) (PAA) thin films, has increasing thermal conductivity as the strength of ionization increases and can reach a thermal conductivity up to ~1.2 W/m.K.[11] It was proposed that due to the Coulombic repulsion between neighboring monomers along the backbone, the polymer chains are stretched, which increases the chain extension and leads to higher thermal conductivity.[11]

However, according to the counterion condensation theory, in dilute polyelectrolyte solutions the polymer chain will collapse when the electrostatic interaction is strong, and counterions will condense on the polymer chain.[16-21] While condensed polyelectrolytes can be different from the dilute solutions, studies also show that the polymer $R_g$ decreases when either the polymer or the salt concentrations increase in the solution.[22-24] When the polyelectrolytes and counterion concentrations increase, the monomer-counterions attraction increases, which enhances counterion condensation and undermines the monomer-monomer electrostatic repulsion, leading to polymer chains collapse.[25-30] Therefore, in bulk ionized PAA, where the counterion condensation effect will be strong, the polymer chains should contract, and thus the high thermal conductivity is unlikely from the polymer chain extension effect.[11] Xie et al. studied a series of bulk polyelectrolytes including PAA, and the largest thermal conductivity measured was ~0.67 W/m.K,[31] adding to the uncertainty related to the mechanism of thermal transport enhancement in polyelectrolytes.

Molecular dynamics (MD) simulations have greatly helped the understanding of the fundamentals of thermal transport in polymers.[7-9, 14, 32] The ability to quantitatively characterize the conformation of polymer chains in the amorphous phase has led to the observation of a positive relation between $R_g$ and thermal conductivity.[14, 15] The ability to decompose the total thermal conductivity into contributions from different interatomic interactions (e.g., bonding interaction, van der Waals (vdW) and Coulombic interactions)[14, 15, 33-35] reveals the critical role of thermal transport along the chain covalent backbone and elucidate the reason of the $R_g$ and thermal conductivity relationship.[14, 15] In this study, we use MD simulations to uncover the relationship between the ionization strength, the conformation of polyelectrolyte chains and their thermal conductivity. It is found that the thermal conductivity increases with the level of ionization, and so is the contribution of Coulombic interactions to the total energy of the amorphous PAA. However, the increase in thermal conductivity is not directly contributed by the stronger Coulombic interactions, but instead by the Lennard-Jones (LJ) interactions. It is found that the counterions condensed around the ionized polyelectrolyte functional groups lead to strong attraction between them, which makes the LJ interactions to be largely in the repulsive region. This leads to strong repulsive interatomic forces and thus more efficient thermal energy transfer through such forces.

## 2. MODELS AND METHODS

**2.1. The Polyelectrolyte Model and Simulation Setup.** All the MD simulations are carried out using the Large-scale Atomic/Molecular Massively Parallel Simulator (LAMMPS).[36] A time step of 0.25 fs is used. Periodic boundary conditions (PBC) are applied in all three directions. The Ewald summation method based on particle-particle particle-mesh (PPPM) algorithm is used to calculate the long-range Coulombic interactions with an error parameter of 0.0001.[37-39] The consistent valence force field (CVFF) is used,[40, 41] and the parameters are listed in Section 1 in Supporting Information (SI). Single chains of PAA and its corresponding ionized forms with sodium counterions (Na$^+$) PANa, are built in *BIOVIA Materials Studio*.[42] The single PAA or PANa chains with 50 monomers are firstly relaxed under the NPT ensemble (600 K, 1 atm) until the volume converges. Then the single chain is replicated to 100 chains with 45200 atoms to make a bulk polyelectrolyte. The bulk model is relaxed under NPT (1500 K, 1atm) and NVT (1500 K) until the chain conformation ($R_g$) converges. Then it is cooled to 300 K under NPT (300 K, 1 atm) and NVT (300 K) until the density converges. A final 0.5 ns NVT (300 K) relaxing is applied before the 2 ns NVE productive simulation. The detailed simulation procedure is written in SI Section 2.

**2.2. The Non-Equilibrium Molecular Dynamics (NEMD) and Heat Flux Decomposition Methods.** Figure 1 shows the scheme of using the NEMD method to calculate the thermal conductivity of a polyelectrolyte model. Using the Langevin thermostat,[43] Fig. 1a shows a heat source of ~10 Å in length is applied in the middle and two heat sinks of ~5 Å in length are applied on the edge (the PBC combines two heat sinks into one). When the system reaches steady state, the thermal conductivity is calculated by Fourier's law in Eq. (1). The heat flux ($J$) is calculated from the energy tally recorded on the two thermostats in Eq. (2), Fig. 1b. The temperature gradient ($dT/dx$) is calculated from the temperature distribution, Fig. 1c. The detailed NEMD method is written in SI Section 3.

$$k = \frac{-J}{(dT/dx)} \tag{1}$$

$$J = \frac{1}{2S}\left(\left|\frac{dQ_{in}}{dt}\right| + \left|\frac{dQ_{out}}{dt}\right|\right) \tag{2}$$

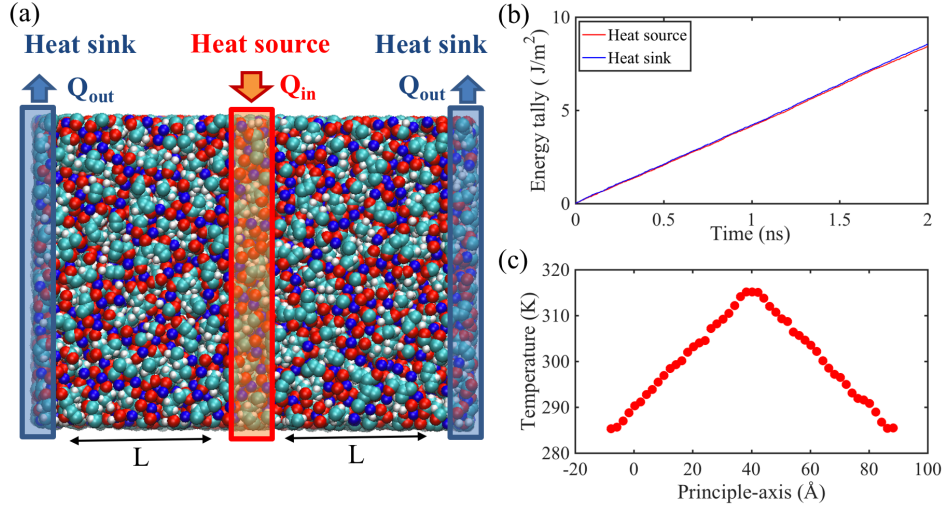

**Figure 1.** The NEMD simulation method for calculating the thermal conductivity: (a) Two thermostats, heat source at 315 K and heat sink at 285 K, are applied to the simulation box. (b) Energy tallies (i.e., cumulative energy extracted from the heat sink or added to the heat source) per cross-sectional area during a 2 ns simulation. Heat flux is taken as the slope of the curves from 1-2 ns. (c) The temperature profile of the polyelectrolyte at steady state. Temperature gradient is taken as the slope of the linear portions of the curves.

At the atomistic level, thermal energy is transferred by atoms doing work to one another through the interatomic interactions. The heat flux along the x-axis ($J_x$) can be written as:

$$J_x = \frac{1}{V}\left\{\sum_i v_{x,i} * e_i + \frac{1}{2}\sum_i \sum_{i<j}\left[\vec{f}_{ij} * (\vec{v}_i + \vec{v}_j) * \vec{r}_{x,ij}\right]\right\} \quad (3)$$

where $V$ is the volume of the box, $v_{x,i}$ is the velocity of atom $i$ along the x-axis, $e_i$ is the kinetic energy of atom $i$, $\vec{f}_{ij}$ is the pairwise interaction force between atoms $i$ and $j$, $\vec{v}_i$ and $\vec{v}_j$ are the velocity vectors of atoms $i$ and $j$, and $\vec{r}_{x,ij}$ is the distance along the x-axis. For simplicity, only two-body terms are shown in Eq. (3), but three- and four-body terms have been included in our decompositions. In LAMMPS the total heat flux is decomposed into energy convection, bonding interaction (bond, angle, torsion, improper torsion), long-range Coulombic interaction (KSpace) and pair interactions (Fig. S2 in SI).[35, 36] By modifying pair interaction code in LAMMPS, we can further separate the pair contribution into short-range Coulombic and LJ contributions (Fig. S3 in SI). In this study, the LJ contribution is further separated into the inter-nuclei repulsive ($4\varepsilon(\frac{\sigma}{r})^{12}$) and vdW attractive ($-4\varepsilon(\frac{\sigma}{r})^6$) portions. In SI Section 4 we provide more information about the heat flux decomposition method.

## 3. RESULTS AND DISCUSSIONS

**3.1. The Thermal Conductivity Comparison Between Our Model And Experimental References.** Figure 2 shows the model calculated thermal conductivity values of PAA at different ionization ratios. As the ionization strength increases from 0% to 100%, the thermal conductivity increases monotonically from ~0.30 to ~0.67 W/m.K, which agrees well with experimental measurements from Ref [11]. The calculated thermal conductivity (~0.30 W/m.K) of the un-ionized PAA agrees with experimental values of 0.28-0.34 W/m.K,[11] and ~0.37 W/m.K[44] (see Fig. 2 reference data points). It was experimentally measured that blade-coated thick (1.5-5.5 μm) PAA ionized films prepared with pH~12 had a thermal conductivity of ~0.62 W/m.K.[11] It was also measured that at pH~12, the ionization ratio was around 90%,[11] at which our MD simulations predicted a thermal conductivity of ~0.59 W/m.K, agreeing well with the experimental measurement. However, in another study, a 100 nm-thick spin-coated ionized PANa film was measured to have thermal conductivity of ~0.45±0.03 W/m.K,[31] which was much lower than that measured from Ref. [11]. One potential difference in these experiments is the ratios of ionization. The PANa measured in Ref. [31] was from Sigma-Aldrich which labeled a pH of 6-9, and this range would correspond to an ionization range of 56-90% according to Ref.

[11]. In this range, our predicted thermal conductivity is 0.45-0.59 W/m.K, with the lower bound agreeing exactly with the measured data from Ref. [31]. We also examine the density of the PAA with different ionization strengths and find a monotonic increase as the ionization ratio increases (Fig. S4 in SI). The densities are predicted to be 1.34 g/cm$^3$ for PAA and 1.76 g/cm$^3$ for PANa, respectively, which are similar to the experimental values of 1.22-1.41 g/cm$^3$ for PAA[11, 45, 46] and 1.70 g/cm$^3$ for PANa.[46] These indicate that our MD simulation predict both morphology and thermal conductivity reasonably. It is worth noting that although density increases, the $R_g$ of polymer chains are not decreasing as ionization increases. Actually, the density increase is mainly due to the addition of heavier Na$^+$ counterions instead of the shrinkage of polymer chains.

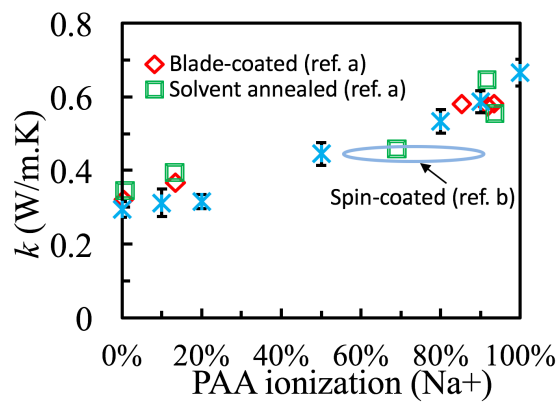

**Figure 2.** Calculated thermal conductivity of PAA with different ionization ratios and its comparison with experimental measurements from a – Ref. [11] and b – Ref. [31].

**3.2. The Molecular Conformation Has No Effect On Thermal Conductivity.** It has been previously revealed that polymer thermal conductivity can be closely related to its chain conformation.[3, 15, 47] In amorphous non-ionized polymers, the thermal conductivity increases when spatial extension of the polymer chains ($R_g$) increases. This is because thermal transport in amorphous polymer is dominated by the covalent bonds along the chain backbones and when such backbones are extended, thermal conductivity increases.[14, 15] In bulk condensed polyelectrolytes, where the counterion concentration is high and the interchain interaction is strong, the counterion condensation theory suggests that the polymer chain will collapse.[16-18] Thus, the observed thermal conductivity increase (in Fig. 2) as ionization increases is unlikely from the chain extension in the polyelectrolyte. Figure 3a shows the conformation of three chains respectively out of the relaxed amorphous PAA structures with ionization ratios of 0%, 50% and 100%. Visually, there is no significant difference in the spatial extension of these chains. We quantitatively characterize the polymer chains of PAA with different ionization ratios by calculating $R_g$, but the chains are always collapsed (Fig. 3b). This finding is inconsistent with the proposed mechanism to explain the enhanced thermal conductivity of PAA when ionization increases,[11] which hypothesized that the Coulombic repulsion of the ionized monomers straightens the chains and thus increases heat transfer along the chain backbone.[11] We believe that the increasing thermal conductivity of ionized PAA is not due to the polymer chain stretching. While we can not claim that our in silico chain conformation is the same as that in experiments,[11] by strengthen the PANa backbone we can increase the $R_g$ from 12.3 to 16.1 Å in the model, but the thermal conductivity merely increases from ~ 0.67 to ~ 0.70 W/m.K.

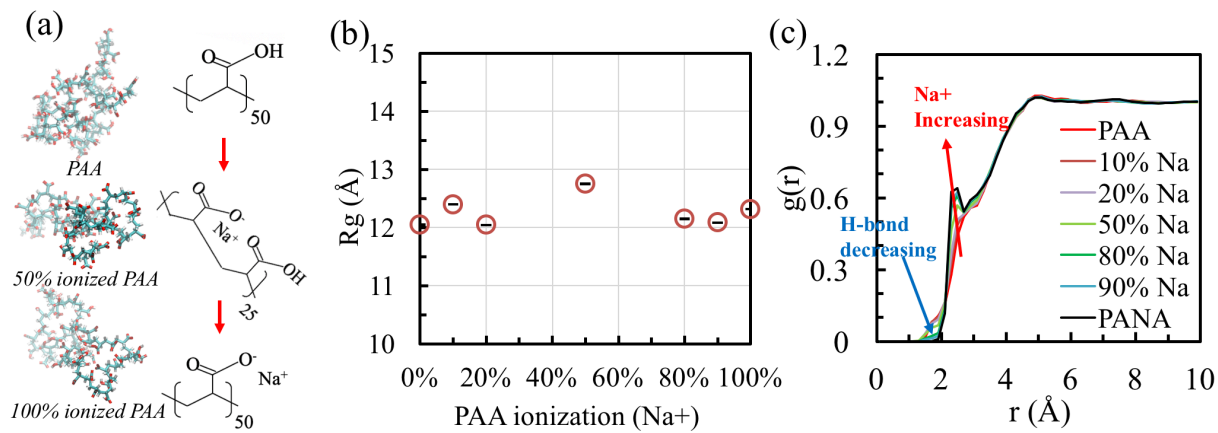

**Figure 3.** (a) Relaxed PAA chains in the amorphous state with different ionization ratios of 0%, 50% and 100%. (b) The $R_g$ of PAA chains at different ionization ratios. (c) RDF involving all atoms except those separated by less than three covalent bonds.

**3.3. The Effect Of Additional Cations On Thermal Conductivity.** We also characterize the morphology by calculating the radial distribution function (RDF) involving all atoms except those separated by less than three covalent bonds. The RDF shows two distinct characters when the ionization ratio increases (Fig. 3c). Firstly, a shoulder peak at ~1.9 Å decreases, which is related to the decreasing number of hydrogen bond formed between the -COOH groups [48] due to the deprotonation process. Secondly, a sharp peak at ~2.5 Å increases, which is due to more $Na^+$ are condensed to the deprotonated $-COO^-$ group. Additional RDF plots (Fig. S5 in SI) involving only the O, H atoms on -COOH, and the $Na^+$ atoms confirm these observations. Another important observation from Fig. 3c is that except the two peaks related to hydrogen bonds and $Na^+$ counterions, the rest of the profile stay unchanged, which is consistent to the almost unchanged $R_g$ of the polymer chains shown in Fig. 3b.

From the above analyses, it can be implied that the increasing thermal conductivity is related to the addition of the $Na^+$ counterions, which can contribute to stronger Coulombic interactions.

Indeed, the amplitude of potential energies due to Coulombic interaction monotonically increase as the ionization increases (Fig. 4a). The negative sign of the Coulombic potential is due to the interaction between the ionized -COO$^-$ group and the counterion Na$^+$. Meanwhile, the LJ potential energy, which is much smaller in amplitude than the Coulombic potential, increases up to a positive value, indicating that the interatomic distances are shifting to the repulsive regime.

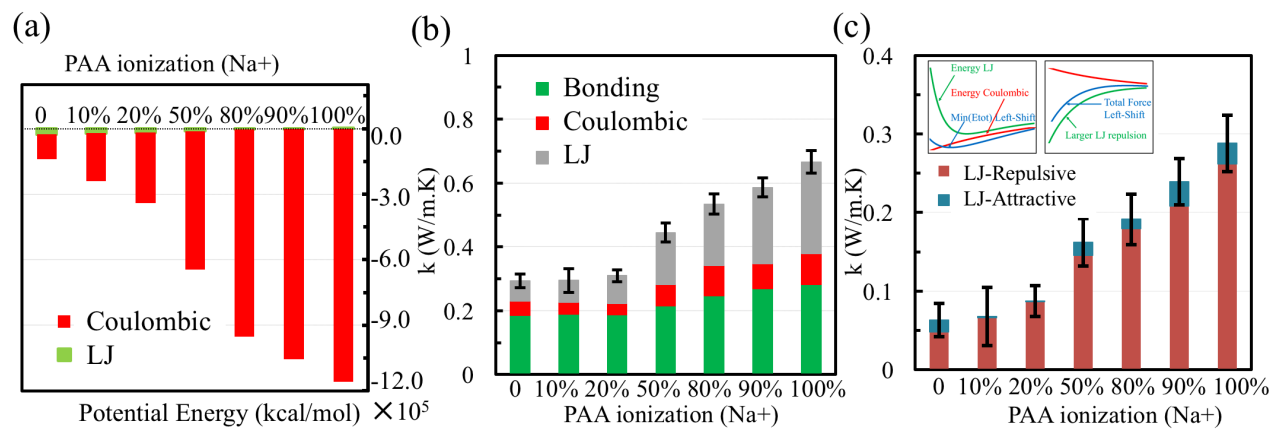

**Figure 4.** (a) Coulombic and LJ potential energies. (b) Thermal conductivity contributions from bonding, Coulombic and LJ interactions. (c) Thermal conductivity decomposition of the LJ part into the LJ repulsive and attractive interaction contributions. The left inserted scheme of Coulombic and LJ potential energy as a function of interatomic distance indicates that the Coulombic potential brings the total energy to another minimum at the LJ repulsive potential part. The right inserted scheme is the Coulombic and LJ forces (derivatives of the potential curves), which shows LJ repulsive forces can dominate the forces in short interatomic distance ranges.

**3.4. The Thermal Conductivity Decomposition.** We further decompose the total thermal conductivity into contributions from bonding, Coulombic and LJ interactions as shown in Fig. 4b. Surprisingly, the increase in thermal conductivity is not mainly due to the increase in the Coulombic contribution but instead from the LJ part. Moreover, the LJ-repulsive contribution is dominant over the LJ-attractive contribution. Figure 4a shows that the Coulombic potential significantly increases with the ionization ratio, and thus it is counter-intuitive that the Coulombic contribution to thermal conductivity only increase slightly. The bonding contribution also increases slightly, which is related to the fact that C-C bond length in PANa is slightly compressed compared to that of PAA (Fig. S6).

Since Na+ atoms replaced the light hydrogen atoms in the -COOH group after ionization, the atomic velocity should have decreased. As a result, to increase heat flux, the interatomic forces should have increased. We analyze the statistics of the amplitude of all the LJ forces, and the histogram in Fig. 5 indeed show that there are more LJ forces with amplitude >10 kcal/mol.Å as ionization increases. We further analyze the histograms of LJ forces on different types of atoms including O, Na, H and C (Fig. 5, green arrows). It can be seen that the higher population of stronger forces (>10 kcal/mol.Å) are from those on the O and Na atoms. Further decomposition of the LJ forces on the O atoms shows that the O-Na LJ forces are dominant (Fig. 5, red arrows). Therefore, ionization leads to stronger LJ forces between the ionized groups and the counterions. On the other hand, while there are stronger Coulombic forces emerging as ionization increases (Fig. S7), both the amplitude of the forces and population increases are less significant than their LJ counterparts. As previously noted in Fig. 4a, the LJ interaction involving Na+ atoms should be largely in the repulsive region, which scales with $r^{-13}$, while that of the Coulombic interaction only scales with $r^{-2}$ (see schematic in the insets of Fig. 4c). It is thus understandable that LJ provide stronger forces for the same atomic pairs. We further decompose the LJ contribution to thermal conductivity into the inter-nuclei repulsive ($4\varepsilon(\frac{\sigma}{r})^{12}$) and vdW attractive $(-4\varepsilon(\frac{\sigma}{r})^6)$ portions. It is confirmed that the repulsive portion is dominant in the overall LJ contribution to thermal conductivity (Fig. 4c).

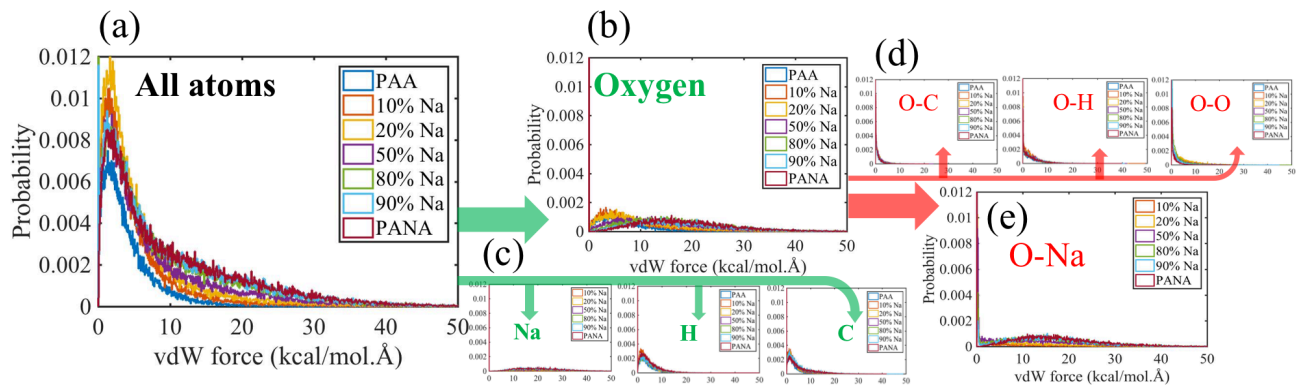

**Figure 5.** Histogram of the amplitude of all LJ forces on (a) all atoms, and the decomposition (green arrows) into the LJ forces on the (b) O, and (c) $Na^+$, H and C atoms. The LJ forces on O is further decomposed (red arrows) into the forces between (d) O-O, O-C, O-H and (e) O-Na pairs.

**3.5. The Partial Charge Scaling Effect On Thermal Conductivity.** We further perform a parametric study on the 100% ionized PANa by scaling the partial charges on the atoms from 50% of the original values to up to 200% (See Section 6 in SI). We find that the thermal conductivity of PANa increases monotonically as the Coulombic interaction scales up (Fig. 6a). By decomposing the thermal conductivity, we find that the thermal conductivity increase from the LJ contribution is more than the Coulombic and Bonding contributions (Fig. 6a). Figure 6b shows that when the partial charge scales up, both the $Na^+$ peak at ~2.5 Å and the shoulder peak at ~4 Å related to the bulk density are left shifted. Due to the stronger Coulombic interactions, the LJ forces largely shift to the repulsive regime and the magnitude of the LJ force increases (Fig. S10 in SI), and thus the heat transfer through LJ interaction is enhanced. It turns out that the strong LJ repulsive interaction is triggered by the Coulombic interaction enhancement and contributes the most in thermal transport.

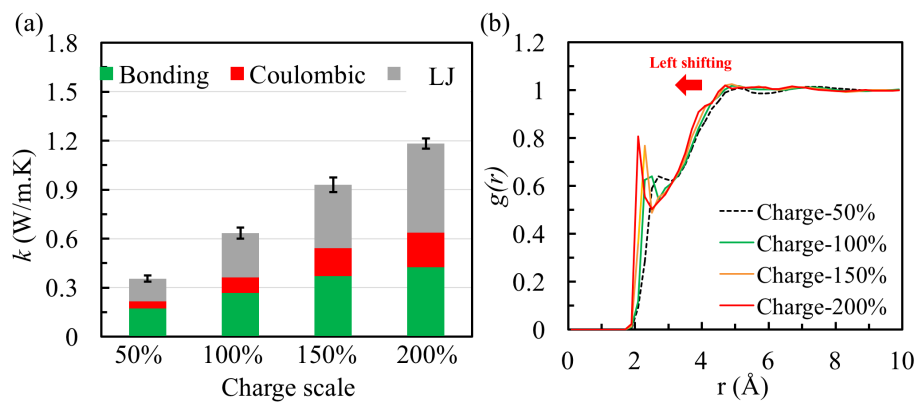

**Figure 6.** The partial charge scaling effects on: (a) thermal conductivity contributions from bonding, Coulombic and LJ interactions, (b) RDF plots of all the atoms (exclude bond, angle and dihedral related atoms) in PANa.

**3.6. The Pressure Scaling Effect On Thermal Conductivity.** A recent study shows that ionized PANa thin films (10-35 nm thick) can reach a thermal conductivity ~1.2 W/m.K in stark contrast to that of the micro-films (~0.62 W/m.K).[11] The thermal conductivity of polymer thin films can indeed be thickness-dependent, which may be related to their density.[49, 50] For example, when the poly(methyl methacrylate) (PMMA) film thickness decreases, its density decreases[50] and so does its thermal conductivity.[49] Some polymer, however, shows an increase in density when the thin film thickness decreases, such as polystyrene.[50] We simulate the density effect by compressing the simulation cell in MD with high pressure up to ~11.2 GPa (See Section 7 in SI). We find that the thermal conductivity of the 100% ionized PANa increases from 0.65 to 1.09 W/m.K after the pressure of ~11.2 GPa is applied (Fig. 7a). This is similar to an experimental result for amorphous PMMA, which shows that the thermal conductivity increases from 0.2 to 0.5 W/m.K when a pressure 12 GPa is applied.[51, 52] By decomposing the thermal conductivity, we find that the LJ interaction still contributes dominantly to the thermal conductivity increase (Fig. 7a). The RDF plot of PANa under different pressure shows that when the pressure increases, the Na$^+$ peak at ~2.5 Å increases and the peak at ~4 Å is left shifted, which corresponds to the LJ interaction increasing (Fig. 7b). Although the high pressure increases the density from ~1.8 to ~2.2 g/cm$^3$ (Fig. S11), we believe this is not the dominant effect, when we compare the normalized thermal conductivity increasing and density increasing (Fig. S12). By pressurizing the polymer, LJ interaction shifts to the repulsive regime and the stronger LJ forces are more populated (Fig. S13), the speed of sound is also enhanced (Table S1). As a result, there is indeed a possibility to achieve higher thermal conductivity in thinner films if the density can be increased significantly.

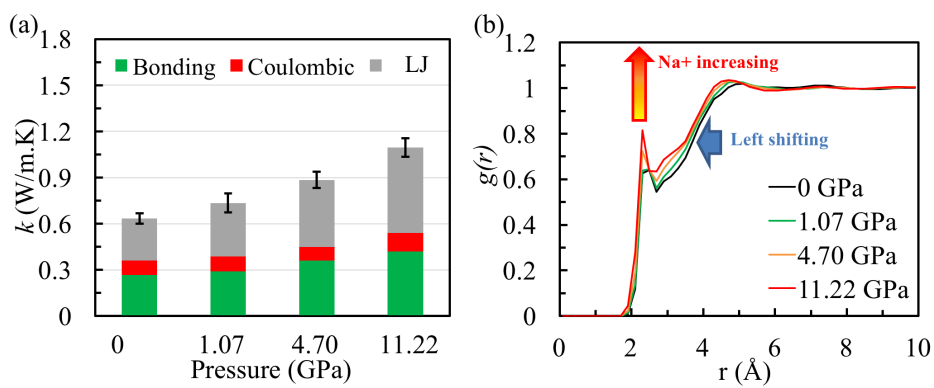

**Figure 7.** The pressure effects on: (a) thermal conductivity contributions from bonding, Coulombic and LJ interactions, (b) RDF plots of all the atoms (exclude bond, angle and dihedral related atoms) in PANa.

## 4. CONCLUSIONS

In summary, we use MD simulations to show that the thermal conductivity of bulk amorphous PAA increases monotonically when the ionization strength increases. We find that counterions, which condense around the ionized PAA side groups, interact strongly with the ionized -COO$^-$ group via Coulombic interaction. As a result, the polymer chain stays collapsed and the thermal transport through bonding interaction does not increase. While the Coulombic interaction drags atoms closer, which induces strong repulsive LJ forces between the ions and the ionized groups. It is such repulsive LJ forces that contribute significantly to the increase in thermal conductivity. Using the heat flux decomposition method, we find that the thermal conductivity increasing by ionization is dominated by the LJ interaction and specifically the LJ-repulsive interaction. The Coulombic interaction does not contribute to the thermal transport directly, but it enhances thermal transport through the LJ interaction. By scaling up the partial charge in PANa, the thermal conductivity from bonding, Coulombic and LJ contributions all increases, but the LJ contribution is dominant. We also find that the thermal transport through LJ interaction can further be triggered by high pressures and the thermal conductivity of PANa at 11.2 GPa reaches 1.09 W/m.K. Reducing the inter-atomic distance helps the LJ force shifts to the strong repulsive regime and increase the LJ force magnitude, which in turn increases thermal conductivity.

## ACKNOWLEDGMENTS

The authors acknowledge the financial support from DuPont Young Professor Award and the Dorini Family endowed professorship. This computation was supported in part by the University of Notre Dame, Center for Research Computing, and NSF through XSEDE resources provided by TACC Stampede-II under grant number TG-CTS100078.## AUTHOR INFORMATION

**Corresponding Author**

E-mail: tluo@nd.edu

**ORCID**

Xingfei Wei: 0000-0001-5924-1579

Tengfei Luo: 0000-0003-3940-8786

**Notes**

The authors declare no competing financial interest.
## ASSOCIATED CONTENT

**Supporting Information**

The Supporting Information (SI) includes the following content: simulation model parameters, heat flux decomposition method, ionization effect on thermal conductivity, Coulombic interaction scaling effect on thermal conductivity, pressure effect on thermal conductivity and a few water molecule effect on thermal conductivity.

# REFERENCES

(1) http://www.engineeringtoolbox.com/thermal-conductivity-d_429.html.

(2) Henry, A. Thermal Transport in Polymers. *Annu. Rev. Heat Transfer* **2013**, *17*.

(3) Luo, T.; Chen, G. Nanoscale Heat Transfer–from Computation to Experiment. *Phys. Chem. Chem. Phys.* **2013**, *15*, 3389-3412.

(4) Chen, H.; Ginzburg, V. V.; Yang, J.; Yang, Y.; Liu, W.; Huang, Y.; Du, L.; Chen, B. Thermal Conductivity of Polymer-Based Composites: Fundamentals and Applications. *Progress in Polymer Science* **2016**, *59*, 41-85.

(5) Lin, W.; Zhang, R.; Wong, C. Modeling of Thermal Conductivity of Graphite Nanosheet Composites. *J Electron Mater* **2010**, *39*, 268-272.

(6) Shen, S.; Henry, A.; Tong, J.; Zheng, R.; Chen, G. Polyethylene Nanofibres with very High Thermal Conductivities. *Nat. Nanotechnol.* **2010**, *5*, 251-255.

(7) Zhang, T.; Luo, T. Morphology-Influenced Thermal Conductivity of Polyethylene Single Chains and Crystalline Fibers. *J. Appl. Phys.* **2012**, *112*, 094304.

(8) Zhang, T.; Wu, X.; Luo, T. Polymer Nanofibers with Outstanding Thermal Conductivity and Thermal Stability: Fundamental Linkage between Molecular Characteristics and Macroscopic Thermal Properties. *J. Phys. Chem. C* **2014**, *118*, 21148-21159.

(9) Henry, A.; Chen, G. High Thermal Conductivity of Single Polyethylene Chains using Molecular Dynamics Simulations. *Phys. Rev. Lett.* **2008**, *101*, 235502.

(10) Kim, G.; Lee, D.; Shanker, A.; Shao, L.; Kwon, M. S.; Gidley, D.; Kim, J.; Pipe, K. P. High Thermal Conductivity in Amorphous Polymer Blends by Engineered Interchain Interactions. *Nature materials* **2015**, *14*, 295-300.

(11) Shanker, A.; Li, C.; Kim, G.; Gidley, D.; Pipe, K. P.; Kim, J. High Thermal Conductivity in Electrostatically Engineered Amorphous Polymers. *Science Advances* **2017**, *3*, e1700342.

(12) Xu, Y.; Wang, X.; Zhou, J.; Song, B.; Jiang, Z.; Lee, E. M.; Huberman, S.; Gleason, K. K.; Chen, G. Molecular Engineered Conjugated Polymer with High Thermal Conductivity. *Science advances* **2018**, *4*, eaar3031.

(13) Singh, V.; Bougher, T. L.; Weathers, A.; Cai, Y.; Bi, K.; Pettes, M. T.; McMenamin, S. A.; Lv, W.; Resler, D. P.; Gattuso, T. R. High Thermal Conductivity of Chain-Oriented Amorphous Polythiophene. *Nat. Nanotechnol.* **2014**, *9*, 384-390.

(14) Zhang, T.; Luo, T. Role of Chain Morphology and Stiffness in Thermal Conductivity of Amorphous Polymers. *J. Phys. Chem. B* **2016**, *120*, 803-812.

(15) Wei, X.; Zhang, T.; Luo, T. Chain Conformation-Dependent Thermal Conductivity of Amorphous Polymer Blends: The Impact of Inter-and Intra-Chain Interactions. *Phys. Chem. Chem. Phys.* **2016**, *18*, 32146-32154.

(16) Manning, G. S. Limiting Laws and Counterion Condensation in Polyelectrolyte Solutions I. Colligative Properties. *J. Chem. Phys.* **1969**, *51*, 924-933.

(17) Grønbech-Jensen, N.; Mashl, R. J.; Bruinsma, R. F.; Gelbart, W. M. Counterion-Induced Attraction between Rigid Polyelectrolytes. *Phys. Rev. Lett.* **1997**, *78*, 2477.

(18) Brilliantov, N.; Kuznetsov, D.; Klein, R. Chain Collapse and Counterion Condensation in Dilute Polyelectrolyte Solutions. *Phys. Rev. Lett.* **1998**, *81*, 1433.

(19) Deserno, M. *Counterion condensation for rigid linear polyelectrolytes* **2000**.

(20) Morawetz, H. Revisiting some Phenomena in Polyelectrolyte Solutions. *Journal of Polymer Science Part B: Polymer Physics* **2002**, *40*, 1080-1086.

(21) Gavrilov, A.; Chertovich, A.; Kramarenko, E. Y. Dissipative Particle Dynamics for Systems with High Density of Charges: Implementation of Electrostatic Interactions. *J. Chem. Phys.* **2016**, *145*, 174101.

(22) Liu, S.; Ghosh, K.; Muthukumar, M. Polyelectrolyte Solutions with Added Salt: A Simulation Study. *J. Chem. Phys.* **2003**, *119*, 1813-1823.

(23) Reed, W. F.; Ghosh, S.; Medjahdi, G.; Francois, J. Dependence of Polyelectrolyte Apparent Persistence Lengths, Viscosity, and Diffusion on Ionic Strength and Linear Charge Density. *Macromolecules* **1991**, *24*, 6189-6198.

(24) Hsiao, P. Chain Morphology, Swelling Exponent, Persistence Length, Like-Charge Attraction, and Charge Distribution Around a Chain in Polyelectrolyte Solutions: Effects of Salt Concentration and Ion Size Studied by Molecular Dynamics Simulations. *Macromolecules* **2006**, *39*, 7125-7137.

(25) Dobrynin, A. V.; Rubinstein, M. Theory of Polyelectrolytes in Solutions and at Surfaces. *Progress in Polymer Science* **2005**, *30*, 1049-1118.

(26) Liao, Q.; Dobrynin, A. V.; Rubinstein, M. Molecular Dynamics Simulations of Polyelectrolyte Solutions: Nonuniform Stretching of Chains and Scaling Behavior. *Macromolecules* **2003**, *36*, 3386-3398.

(27) Liao, Q.; Dobrynin, A. V.; Rubinstein, M. Molecular Dynamics Simulations of Polyelectrolyte Solutions: Osmotic Coefficient and Counterion Condensation. *Macromolecules* **2003**, *36*, 3399-3410.

(28) Liao, Q.; Dobrynin, A. V.; Rubinstein, M. Counterion-Correlation-Induced Attraction and Necklace Formation in Polyelectrolyte Solutions: Theory and Simulations. *Macromolecules* **2006**, *39*, 1920-1938.

(29) Stevens, M. J.; Kremer, K. Structure of Salt-Free Linear Polyelectrolytes. *Phys. Rev. Lett.* **1993**, *71*, 2228.

(30) Stevens, M. J.; Kremer, K. The Nature of Flexible Linear Polyelectrolytes in Salt Free Solution: A Molecular Dynamics Study. *J. Chem. Phys.* **1995**, *103*, 1669-1690.

(31) Xie, X.; Yang, K.; Li, D.; Tsai, T.; Shin, J.; Braun, P. V.; Cahill, D. G. High and Low Thermal Conductivity of Amorphous Macromolecules. *Physical Review B* **2017**, *95*, 035406.

(32) Henry, A.; Chen, G.; Plimpton, S. J.; Thompson, A. 1D-to-3D Transition of Phonon Heat Conduction in Polyethylene using Molecular Dynamics Simulations. *Phys. Rev. B: Condens. Matter Mater. Phys.* **2010**, *82*, 144308.

(33) Ohara, T.; Chia Yuan, T.; Torii, D.; Kikugawa, G.; Kosugi, N. Heat Conduction in Chain Polymer Liquids: Molecular Dynamics Study on the Contributions of Inter-and Intramolecular Energy Transfer. *J. Chem. Phys.* **2011**, *135*, 034507.

(34) Matsubara, H.; Kikugawa, G.; Ishikiriyama, M.; Yamashita, S.; Ohara, T. Equivalence of the EMD- and NEMD-Based Decomposition of Thermal Conductivity into Microscopic Building Blocks. *J. Chem. Phys.* **2017**, *147*, 114104.

(35) Matsubara, H.; Kikugawa, G.; Bessho, T.; Yamashita, S.; Ohara, T. Molecular Dynamics Study on the Role of Hydroxyl Groups in Heat Conduction in Liquid Alcohols. *Int. J. Heat Mass Transfer* **2017**, *108*, 749-759.

(36) Plimpton, S.; Crozier, P.; Thompson, A. LAMMPS-Large-Scale Atomic/Molecular Massively Parallel Simulator, http://Lammps.Sandia.Gov. *Sandia National Laboratories* **2007**, *18*.

(37) Hockney, R. W.; Eastwood, J. W. In *Computer simulation using particles;* crc Press: 1988; .

(38) Kolafa, J.; Perram, J. W. Cutoff Errors in the Ewald Summation Formulae for Point Charge Systems. *Molecular Simulation* **1992**, *9*, 351-368.

(39) Allen, M. P.; Tildesley, D. J. In *Computer simulation of liquids;* Oxford university press: 2017; .

(40) Cranford, S. W.; Buehler, M. J. Variation of Weak Polyelectrolyte Persistence Length through an Electrostatic Contour Length. *Macromolecules* **2012**, *45*, 8067-8082.

(41) Dauber-Osguthorpe, P.; Roberts, V. A.; Osguthorpe, D. J.; Wolff, J.; Genest, M.; Hagler, A. T. Structure and Energetics of Ligand Binding to Proteins: Escherichia Coli Dihydrofolate Reductase-trimethoprim, a Drug-receptor System. *Proteins: Structure, Function, and Bioinformatics* **1988**, *4*, 31-47.

(42) Anonymous*Dassault Systèmes BIOVIA, Materials Studio, Version 8, San Diego: Dassault Systèmes, 2014*.

(43) Schneider, T.; Stoll, E. Molecular-Dynamics Study of a Three-Dimensional One-Component Model for Distortive Phase Transitions. *Physical Review B* **1978**, *17*, 1302.

(44) Xie, X.; Li, D.; Tsai, T.; Liu, J.; Braun, P. V.; Cahill, D. G. Thermal Conductivity, Heat Capacity, and Elastic Constants of Water-Soluble Polymers and Polymer Blends. *Macromolecules* **2016**, *49*, 972–978.

(45) Mark, J. E. In *Physical properties of polymers handbook;* Springer: 2007; Vol. 1076.

(46) Hiraoka, K.; Shin, H.; Yokoyama, T. Density Measurements of Poly (Acrylic Acid) Sodium Salts. *Polymer Bulletin* **1982**, *8*, 303-309.

(47) Zhang, T.; Luo, T. High-Contrast, Reversible Thermal Conductivity Regulation Utilizing the Phase Transition of Polyethylene Nanofibers. *ACS nano* **2013**, *7*, 7592-7600.

(48) Visakh, P.; Bayraktar, O.; Picó, G. A. In *Polyelectrolytes: thermodynamics and rheology;* Springer: 2014.

(49) Losego, M. D.; Moh, L.; Arpin, K. A.; Cahill, D. G.; Braun, P. V. Interfacial Thermal Conductance in Spun-Cast Polymer Films and Polymer Brushes. *Appl. Phys. Lett.* **2010**, *97*, 011908.

(50) Unni, A. B.; Vignaud, G.; Chapel, J.; Giermanska, J.; Bal, J.; Delorme, N.; Beuvier, T.; Thomas, S.; Grohens, Y.; Gibaud, A. Probing the Density Variation of Confined Polymer Thin Films Via Simple Model-Independent Nanoparticle Adsorption. *Macromolecules* **2017**, *50*, 1027-1036.

(51) Assael, M.; Botsios, S.; Gialou, K.; Metaxa, I. Thermal Conductivity of Polymethyl Methacrylate (PMMA) and Borosilicate Crown Glass BK7. *Int. J. Thermophys.* **2005**, *26*, 1595-1605.

(52) Hsieh, W.; Losego, M. D.; Braun, P. V.; Shenogin, S.; Keblinski, P.; Cahill, D. G. Testing the Minimum Thermal Conductivity Model for Amorphous Polymers using High Pressure. *Physical Review B* **2011**, *83*, 174205.